# Dislocation distribution, crystallographic texture evolution and plastic inhomogeneity of Inconel 718 fabricated by laser powder-bed fusion

Jalal Al-Lami[1], Thibaut Dessolier[1], Talha Pirzada[2], Minh-Son Pham[1]

[1]Department of Materials, Imperial College London, South Kensington, London SW7 2AZ, UK

[2]Department of Materials, University of Oxford, Oxford OX1 3PH, UK

**Abstract**

Plastic inhomogeneity, particularly localised strain, is one of the main mechanisms responsible for failures in engineering alloys. This work studied the spatial arrangement and distribution of microstructure (including dislocations and grains) and their influence in the plastic inhomogeneity of Inconel 718 fabricated by additive manufacturing (AM). In particular, the density and spatial distribution of geometrically necessary dislocations (GNDs) and the crystallographic texture components were examined in relation to the spatial plastic deformation. *In-situ* tensile testing coupled with microscopy and electron backscatter diffraction (EBSD) was performed to understand the evolution of the as-built microstructure and its effect on the spatial and temporal deformation behaviour. The bidirectional scanning strategy with no interlayer rotation resulted in highly ordered alternating arrangements of coarse Goss-like {110}<001> textured grains separated by fine Cube-like {100}<001> textured grains. The bidirectional strategy also resulted in an overall high density of dislocations (in particular GNDs) with a highly regular pattern: the fine grains contained a very high GND density while the coarse grains contained less GNDs but they were highly oriented to be perpendicular to the scanning direction. Although this scanning strategy resulted in a dominant Cube texture that is desirable for isotropy, the Cube texture gradually weakened and the undesirable Goss component (which was the second most dominant in the as-built microstructure) increased during plastic deformation. Together with the columnar grain morphology and aligned GND arrangement, such texture evolution caused intense plastic inhomogeneity during deformation. The effect of the scanning strategy was examined by replacing the bidirectional strategy with a chessboard strategy involving a 67° layer rotation. The layer rotation in the chessboard strategy resulted in a lower GND density and a much more random distribution of both crystallographic texture and GNDs, with a dominant Cube component (and much lower Goss texture) that remained stable throughout plastic deformation, reducing the plastic inhomogeneity – this is preferable for AM builds.

## 1. Introduction

Additive manufacturing (AM) offers an unprecedented ability to manufacture components of complex geometry [1,2] while simultaneously tailoring the microstructure on the grain level [3–6]. Owing to the layerwise nature of the process, the printing parameters can be altered at different locations within individual layers to enable a fine-scale spatial engineering of the microstructure and achieve desired mechanical properties at specific locations [7]. However, the reliability of metal AM components is still limited primarily by challenges in fabricating high-performance components [8–10]. While advances in processing techniques have improved the understanding of the formation of defects such as porosity and cracks, and ways to minimise the defect densities [11–13], in-depth understanding of the fine microstructure remains insufficient. In particular, the spatial arrangement and distribution of dislocations, grains and detailed crystallographic texture, and how such fine microstructure affects the local plastic deformation spatially in AM components is still unclear. Obtaining such understanding is crucial because most damage initiates in localised deformed regions.

The influential role of the scanning strategy in tailoring the microstructure during AM processing was previously highlighted. Many studies demonstrated the impact of the scanning strategy on the grain microstructure such as the size, morphology and overall crystallographic texture [5,6,14–21] and its influence on the global deformation behaviour [22–27]. Altering the scanning strategy changes the local heat flux, cooling rate and the ratio of thermal gradient ($G$) to solidification rate ($R$), hence controlling the mode and length scale of the solidification microstructure [28,29]. For example, it was demonstrated that the chessboard scanning strategy with 67° interlayer rotation can often interrupt the epitaxial columnar growth and crystallographic texture in AM components [14]. The crystallographic texture is an influential factor in the plastic deformation behaviour spatially and globally. For example, if the ideally random texture is excluded, the Cube {001}<100> texture component results in the least anisotropy, whereas the Goss {110}<001> component causes the most anisotropy [30]. While the dominant crystallographic orientations in as-built AM microstructure were previously reported (such as a strong <001> fibre developing along the building direction in cubic alloys) [14,18], detailed analysis of texture components and their evolution during plastic deformation have not been investigated. Even if a desirable initial texture was created in the as-built condition, it might evolve under mechanical loading into a less desirable one. Therefore, it is important to understand how the texture evolves during plastic deformation.

Amongst all reported scanning strategies, the chessboard strategy with a 67° layer rotation is often used to fabricate components, and multiple studies reported that this strategy helps in improving mechanical properties. For example, the chessboard strategy was reported to result in lower residual stresses [22–24] and reduced anisotropy [14,19]. Such improvements are often attributed to grain microstructure, such as the grain morphology and crystallographic texture. However, the dislocation density and spatial distribution are also influential in the internal stresses (hence residual stresses) and plastic inhomogeneity (hence the anisotropy). Plastic inhomogeneity is often demonstrated by plastic localisation, which is one of the main mechanisms causing the initiation of cracks. Nevertheless, the role of the scanning strategy in controlling the density and spatial distribution of dislocations, such as geometrically necessary dislocations (GNDs), has not been sufficiently studied. Developing a fundamental understanding of the effect of the scanning strategy on the dislocation alignment and distribution will help in obtaining a better understanding of how the local plastic deformation progresses. This understanding will enable the enhancement of the mechanical properties thanks to the ability of engineering the microstructure on a subgrain level via AM processing.

In this study, a detailed investigation of the spatial arrangement and distribution of microstructure, including grains, crystallographic texture components, and dislocation density and distribution in as-built conditions was performed. Two commonly used scan strategies (bidirectional without layer rotation and chessboard with 67° layer rotation) were used to examine the effect of the scan strategy on the microstructure, in particular the texture, and density and distribution of GNDs. In addition, *in-situ* tensile testing coupled with detailed microstructural characterisation was performed to obtain direct observations of links between the microstructure and plastic inhomogeneity. The *in-situ* tests also provide new insights into the evolution of the microstructure with straining and its impact on the spatial deformation behaviour, including plastic localisation. The study was focused on Inconel 718, which is widely used in the aerospace, automotive and power generation industries [31–33].

## 2. Materials and methods

Inconel 718 cylinders of 16 mm diameter and 150 mm length were fabricated horizontally by laser powder-bed fusion (L-PBF) using a Renishaw AM 250 machine. Gas atomised powder of spherical morphology and a particle size distribution of 15-45 μm was used. All cylinders were printed with a laser power of 200 W, hatch spacing of 0.08 mm, point distance of 0.06 mm and exposure time of 0.0001 s. However, the scanning strategy was varied to alter the microstructure. Individual cylinders were printed using a bidirectional strategy with 0° interlayer rotation or a chessboard strategy with 67° interlayer rotation, which are henceforth referred to as B0 and CB67, respectively. The scanning bi-directions in the B0 strategy were parallel to the longitudinal axis of the cylinder. Samples for microscopy and *in-situ* tensile testing, Figure 1, were extracted by wire electrical discharge machining (EDM).

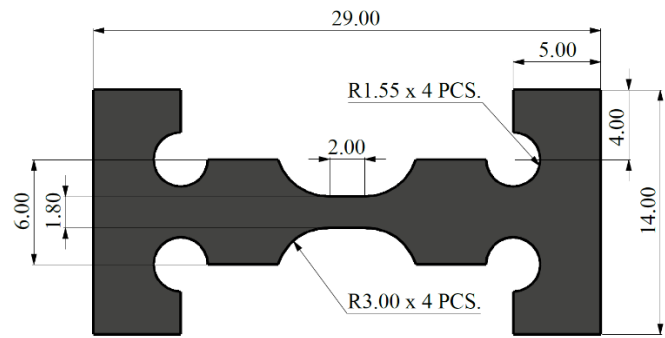

Figure 1. *In-situ* testing sample geometry. All dimensions in mm and figure not to scale.

The samples were prepared by grinding with SiC paper in gradual steps using 800, 1200, 2000 and 4000 P-grades, followed by polishing in a 50% solution of oxide polishing suspension (OPS) mixed with distilled water. To reveal the solidification microstructure, samples were etched with Aqua Regia (freshly mixed HCl with $HNO_3$ in a 3:1 ratio) for 8-10 seconds.

Microstructural characterisation was performed using optical microscopy, scanning electron microscopy (SEM), electron backscatter diffraction (EBSD), and transmission electron microscopy (TEM). Secondary electron (SE) and backscatter electron (BSE) SEM imaging was performed using a voltage of 15 kV in a Zeiss Auriga microscope. Standard EBSD operating condition of 20 kV voltage, 15 mm working distance and 120 μm aperture was used, with a step size of 0.5 μm to capture the crystal orientation spread within the grains at small length scales. A JEOL JEM-2100F microscope operating at 200 kV was used to directly observe dislocations using

TEM. Thin foils for TEM were prepared by twin-jet electropolishing in a Struers TenuPol-5. Electropolishing was performed at 20 kV and -5 °C in a solution of 10% perchloric acid in methanol. Because TEM is not effective in examining the dislocation distribution and density over a large area, EBSD was used to quantitatively characterise the dislocation condition (including density and distribution). It should be noted that EBSD is only able to quantify the geometrically necessary dislocations (GNDs). However, it was found that GNDs co-located with the total dislocations in as-built alloys fabricated by AM [34]. Therefore, the spatial characterisation of GNDs also provides insights into the spatial distribution of total dislocations. In this study, GNDs were measured from the EBSD data using MTEX in MATLAB. The EBSD-GND density calculations are based on the measurement of the crystal orientation spread within individual grains. EBSD was used to quantify the variation in the crystallographic orientation between neighbouring points to calculate the lattice curvature tensor and dislocation density tensor [35]. The mathematical framework on which the MTEX code is based is provided in Pantleon's study [36].

To study the evolution of the microstructure and its correlation with the deformation behaviour, *in-situ* tensile testing experiments using optical microscopy or EBSD mapping were performed separately. In EBSD *in-situ* testing, the samples were held in a Gatan stage fitted inside an FEI Quanta FEG 650 SEM equipped with an EBSD system. EBSD scans were acquired on the undeformed samples and at strain levels of 4%, 7%, 10% and after fracture, using the same EBSD operating condition listed previously but with a step size ranging from 1 to 2 μm. The uniaxial loading was performed on the samples with the loading direction (LD) being parallel to the scanning direction of the B0 strategy. The tests were performed at room temperature and at a strain rate of $10^{-3}$ per second. The evolution of near ideal Cube, Copper, S, Brass and Goss texture components (Table 1) was examined and correlated with the deformation behaviour. Optical microscopy *in-situ* tensile testing was performed by uniaxially loading the samples at the same strain rate of $10^{-3}$ per second in a Gatan stage fitted under an Olympus BX35M optical microscope. One micrograph was captured every second. To correlate the observations with the microstructure, *ex-situ* EBSD maps of the region of interest in *in-situ* optical microscopy tensile testing were acquired before and after deformation. *In-situ* optical microscopy tensile testing helped to better reveal sites of intense topographical changes that cannot be clearly observed under the SEM.

Table 1. Texture components specified in Miller indices and Euler angle notation.

| Texture component | Miller indices | Bunge $(\varphi_1, \theta, \varphi_2)$ |
|---|---|---|
| Cube | $\{001\} <100>$ | $(0, 0, 0)$ |
| Copper | $\{112\} <11\bar{1}>$ | $(40, 65, 26)$ |
| S | $\{123\} <63\bar{4}>$ | $(32, 58, 18)$ |
| Brass | $\{110\} <\bar{1}12>$ | $(35, 45, 0)$ |
| Goss | $\{110\} <001>$ | $(0, 45, 0)$ |

## 3. Results and discussion

### 3.1. Role of scan strategy in grain microstructure and GND distribution

The as-built microstructure of Inconel 718 processed by L-PBF is shown in Figure 2. In both the B0 and CB67 conditions, fine cells (or dendrites) can be seen growing in a direction nearly perpendicular to the melt pool fusion lines, as indicated by the arrows in Figure 2a. This growth direction is anti-parallel to the maximum thermal gradient [14]. However, as the local thermal gradient changes with locations, in particular at the fusion boundary at which there are often changes of 90°, crystal cells (or dendrites) can still epitaxially grow by side-branching, Figure 2b. The same underlying mechanisms of crystal growth were observed in both the B0 and CB67 conditions. However, the scanning strategies induced different microstructure, particularly concerning the grain alignment and crystallographic texture, in agreement with previous reports [14,18,37]. The alignment of melt pools along the vertical building direction in the B0 strategy promotes the continuous epitaxial columnar growth of slender grains along the centreline of consecutive melt pools. On the sides of melt pools, the local changes in thermal gradient promote the epitaxial growth by side-branching with one of the <001> orientations aligned at around 45° to the building direction, making a <101> orientation parallel to the building direction. This results in a microstructure comprised of thin and highly columnar grains of <001> orientation (in the centre of deposition beads/tracks) alternated by coarser grains of <101> orientation at the sides, with respect to the building direction, as shown in Figure 3a. Correspondingly, the view perpendicular to the building direction of samples of the B0 strategy showed a highly ordered arrangement of coarse grains of <101> orientation separated by rows of fine grains of <001> orientation, Figure 3b. Introducing rotation between deposition layers (e.g. by CB67 strategy) disrupts the thermal profile and interrupts the continuous epitaxial growth of cells along the melt pool's centreline observed in the B0 condition, resulting in less columnar grains of more random morphology and size distribution, Figure 3(e-f) and Figure 4. The quantified average grain size along the BD in the B0 and CB67 conditions was 30 μm and 23 μm, respectively. We will show later in this study that the change in the scan strategy alters not only the arrangement, size and distribution of grains, but also the dislocation condition.

TEM examination shown in Figure 2c revealed fine sub-grain cells containing a high dislocation density, particularly at the cell boundaries. Such cells in as-built alloys were reported in other studies [19,38–44] and are believed to be influential to the plastic deformation, in particular the strength.

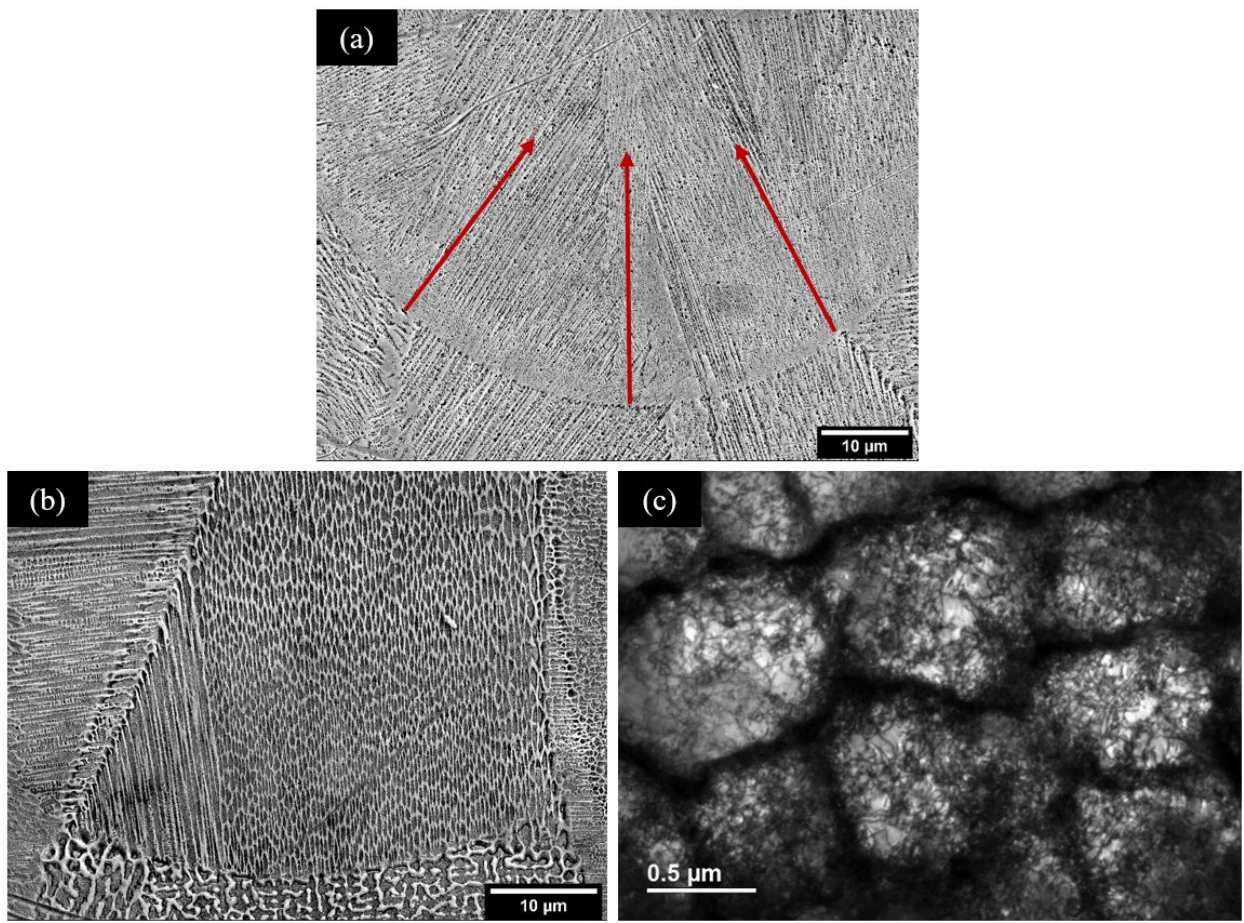

Figure 2. As-built microstructure of Inconel 718 processed by L-PBF. (a) Microstructure in a deposition bead with arrows showing the growth direction of cells that were initially almost perpendicular to the fusion boundary. (b) Side-branching across the fusion boundary was seen in the top left region. (c) Cells containing a high density of dislocations, particularly at the cell boundaries.

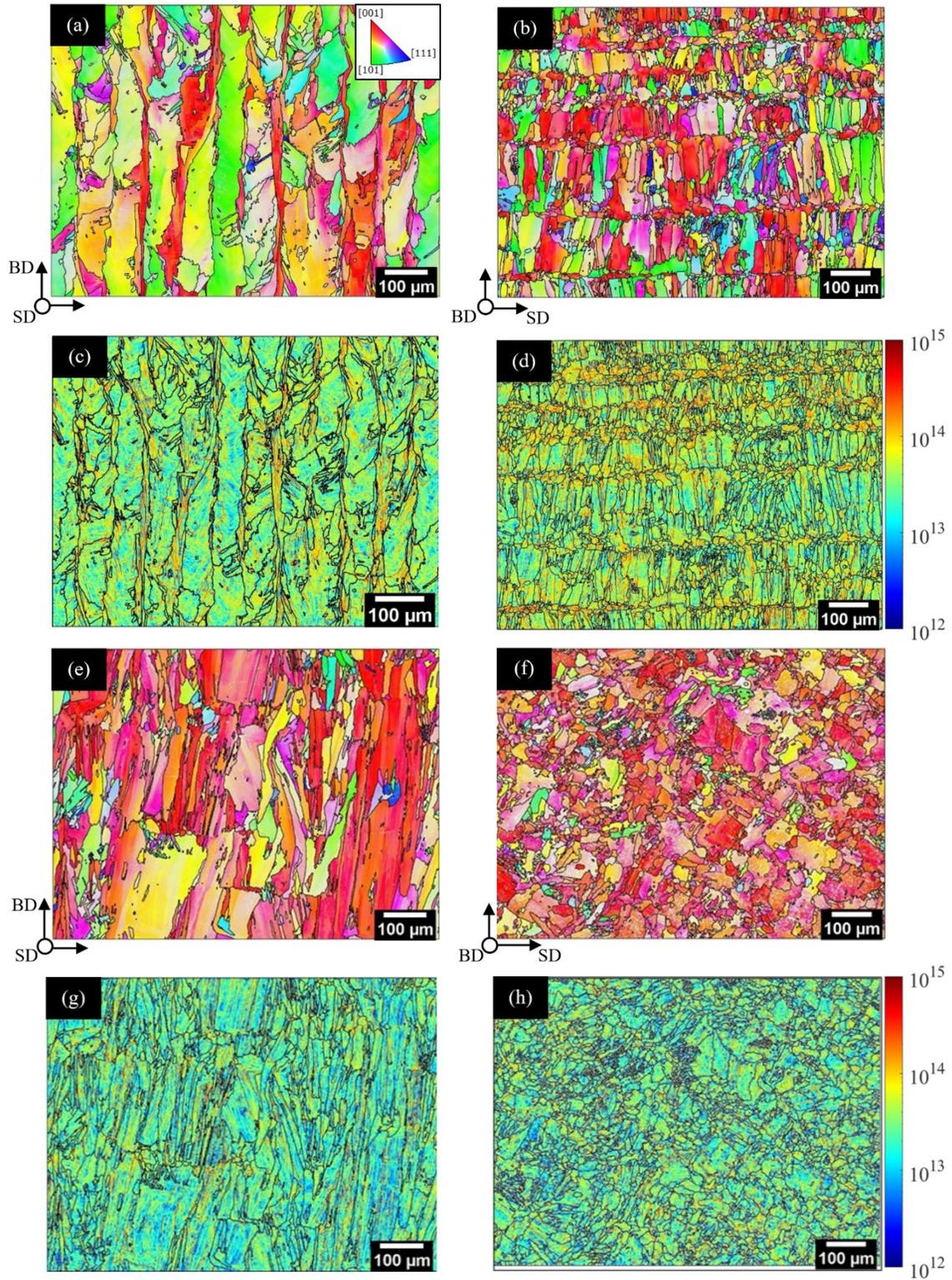

Figure 3. EBSD inverse pole figures (IPF) of the B0 and CB67 conditions with their corresponding GND spatial distribution maps. The B0 condition is presented in (a) to (d), whereas the CB67 is presented in (e) to (h). Note that the left column shows the cross-section along the building direction (BD) with IPF ∥ BD, whereas the right column shows the top-view along the scanning direction (SD) with IPF ∥ SD.

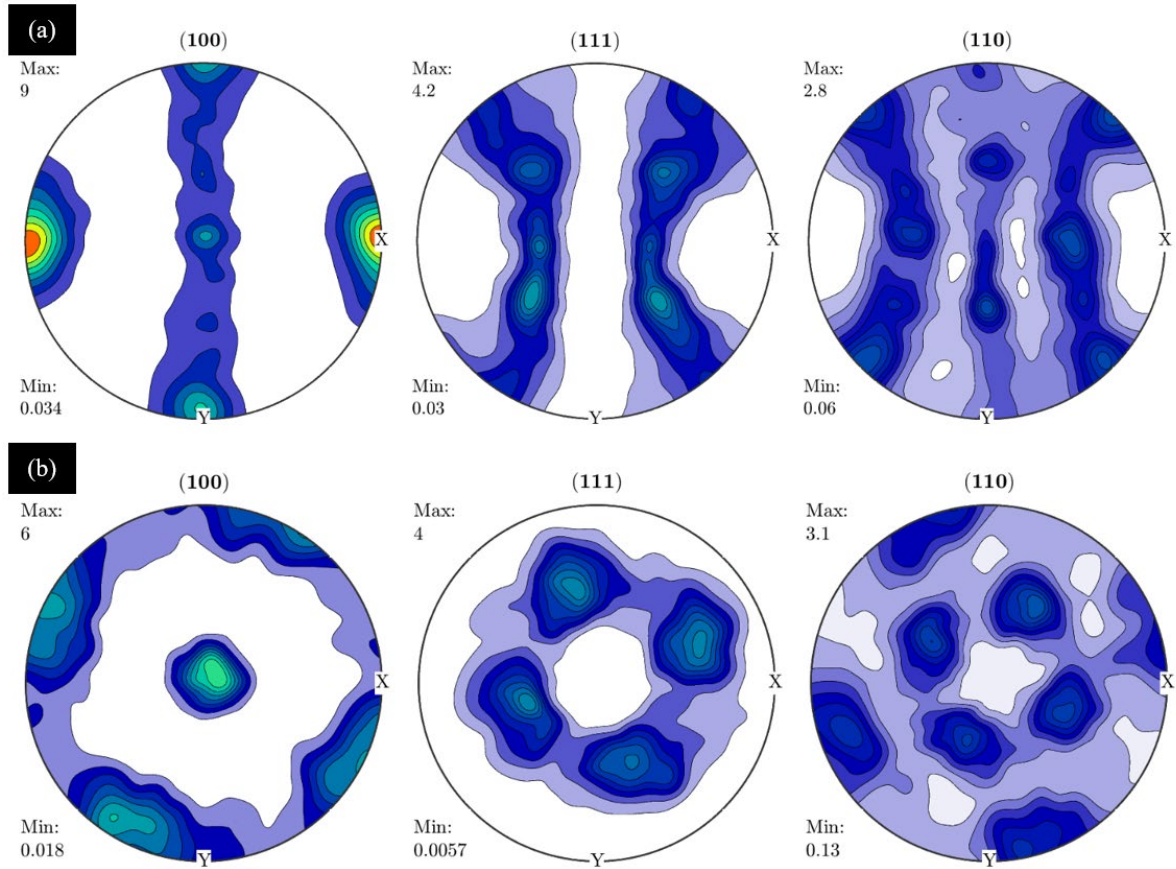

Figure 4. Pole figures of the (a) B0 condition and (b) CB67 condition. Data acquired from the top-view maps. X and Y are perpendicular to the building direction. X is parallel to the longitudinal axis of the samples, which is along the B0 scanning bi-directions, whereas Y is perpendicular to this axis.

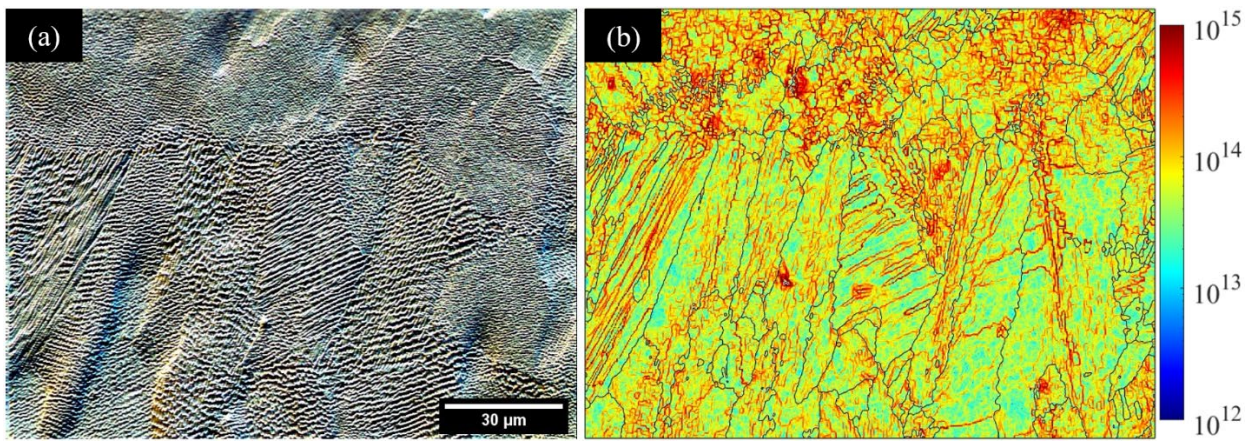

Figure 5. GNDs located at and aligned along the boundaries of cells (or dendrites): (a) SEM micrograph showing the cellular arrays in as-built Inconel 718; (b) corresponding GND spatial distribution map.

The representative spatial distribution of GNDs in the B0 and CB67 conditions is presented in Figure 3(c-d) and (g-h), respectively. The GNDs in the as-built condition are located at and aligned along the boundaries of cells, as shown in Figure 5. This is in agreement with the TEM observation as it showed that most of dislocations were at cell boundaries (Figure 2c). The measurement of GNDs using EBSD enables the correlation of the distribution of GND density with other microstructural features such as the grain size and orientation over areas larger than those captured by TEM.

The top-view GND map of the B0 condition (Figure 3d) shows that the GND alignment is nearly orthogonal to the scanning direction as the cells' growth direction was perpendicular to the scan direction due to the fast laser beam speed. The GND accumulation was particularly very high in the fine grains of the <001> ∥ BD orientation (Figure 3(c-d) and (g-h), and Figure 6). In the B0 condition, the fine grains were along the centreline of deposition tracks (Figure 3c-d). The high GND density in the red <001> elongated centreline grains is believed to have developed primarily because of two reasons. First, although the ordered stacking of melt pools promotes the continuous epitaxial growth (i.e. without side-branching) of the centreline grains, the stacking is not perfect in the process of layer-upon-layer deposition inherent in AM, resulting in a slight misalignment between consecutive layers. This essentially causes some variations in the heat flux along the centreline (hence thermal strain) between the vertical melt pools in consecutive layers. If the variation is not large enough (found to be within about 30° [14]) to interrupt the continuous epitaxial growth, the existing centreline grains would continue growing vertically and dislocations need to be generated to accommodate the thermal strain induced by the variations. Second, the centreline grains are the last region to solidify within a melt pool and their growth is constrained by the surrounding solids that shrink due to thermal contraction during cooling. The surrounding contraction causes tension on the centreline grains, which promotes the generation of dislocations in the fine centreline grains.

Horizontally-printed samples fabricated with the B0 strategy have been reported to have a higher yield stress than samples printed with other strategies and this was previously attributed solely to the grain boundary strengthening effect imposed by the fine grains [15,45]. However, this study demonstrates that the strengthening effect is also due to the higher GND accumulation, particularly in the fine centreline grains induced by this scan strategy. The results also show that the change in scan strategy from B0 to CB67 significantly varied the GND spatial distribution, alignment and density. In the CB67 condition, dislocations were distributed more uniformly, had a more random alignment, and their total density was much lower in comparison with the B0 condition, as shown in Figure 3 and Figure 6. The overall lower density of dislocations in the CB67 condition reduces the global residual stress. Also, the more uniform distribution of dislocations should be beneficial as it reduces the build-up of type-II and type-III residual stresses and the anisotropy of the builds. The reduction in anisotropy enabled by varying the scan strategy helps explain the more isotropic response reported in CB67 builds [14].

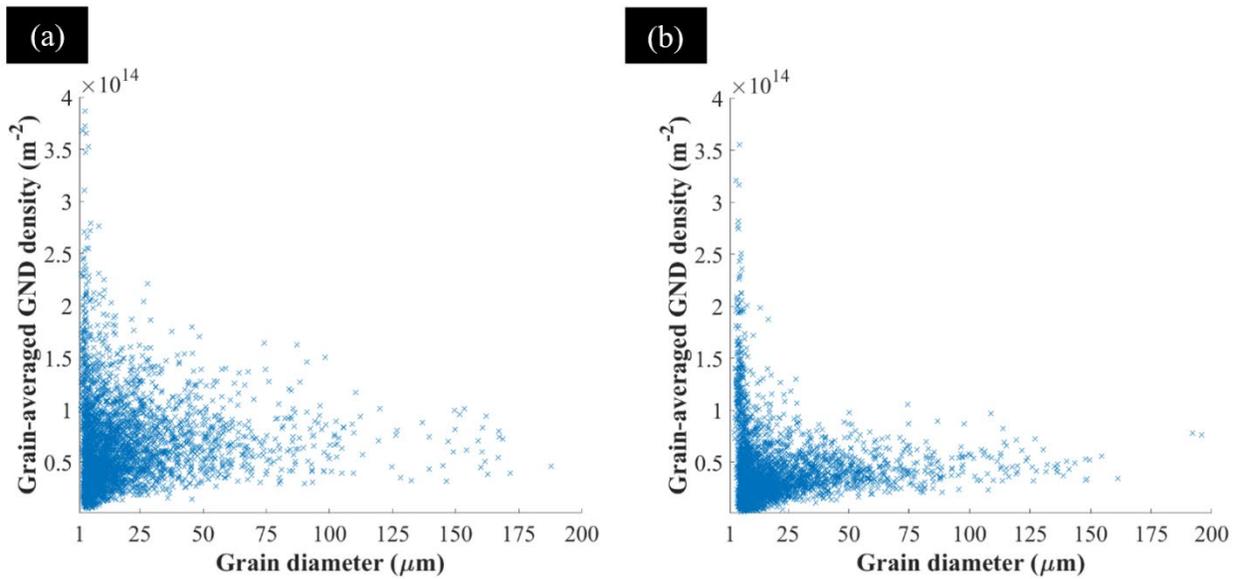

Figure 6. GND density as a function of grain diameter in the as-built and undeformed (a) B0 condition and (b) CB67 condition. The data was measured from the top-view maps.

### 3.2. Microstructure evolution during plastic deformation

The previous section described the as-built microstructure (such as grain microstructure and distribution and density of GNDs), and the effect of the B0 and CB67 scanning strategies on the microstructure. In this section, the influence of the microstructure on the deformation behaviour is examined.

*In-situ* tensile testing in an optical microscope revealed the quick development of inhomogeneous topographical changes. For both the B0 and CB67 conditions, significant variation in surface topography was observed in the vicinity of grain boundaries of significant crystallographic misorientation, e.g. around the cluster of grains labelled "1" in Figure 7a. This surface roughening happened because of plastic strain incompatibility and restricted slip transmission between neighbouring grains of different crystallographic orientations. The difficulty of slip transmission across grain boundaries is governed by the misorientation between the two adjacent grains with respect to the loading direction and the type of grain boundary between them. While the <001> ∥ LD grains were preferably oriented for easy octahedral (111)<110> slip, the <111> ∥ LD grains were not favourably oriented. For slip system activation, <111> ∥ LD grains should rotate to a more favourable orientation, but this rotation is resisted by the neighbouring grains [46]. The difficulties in dislocation slip transmission from the <001> ∥ LD grains to the <111> ∥ LD grains resulted in intense surface topographical change in the vicinity of the grain cluster "1". This effect is further elaborated by another *in-situ* tensile test in SEM/EBSD, Figure 8. The continuity of slip lines across multiple boundaries (Figure 8a-b) suggests good transmission of dislocations between the adjacent soft <001> ∥ LD grains "1" to "2", but such slip lines were discontinued at the boundary between grains "2" and "3" as grain "3" was not preferably oriented for slip and had a high Taylor factor (Figure 8c). Even at a nominal strain of 10%, the slip traces in Figure 8a indicate that most plastic deformation was accommodated by the line of fine <001> ∥ LD grains (that had low Taylor factors), Figure 8c. In comparison to the CB67, plastic inhomogeneity developed much faster and more intensely in the B0 condition. In particular, the localised strain quickly developed in the vicinity of the clustered fine grains with <001> ∥ LD, as shown in the region contained within the dashed yellow rectangles in Figure 7a-b. This line cluster of fine grains was absent in the CB67 microstructure thanks to the layer rotation. The absence of the line of fine grains appeared to be beneficial as it reduced the plastic inhomogeneity in the CB67 condition.

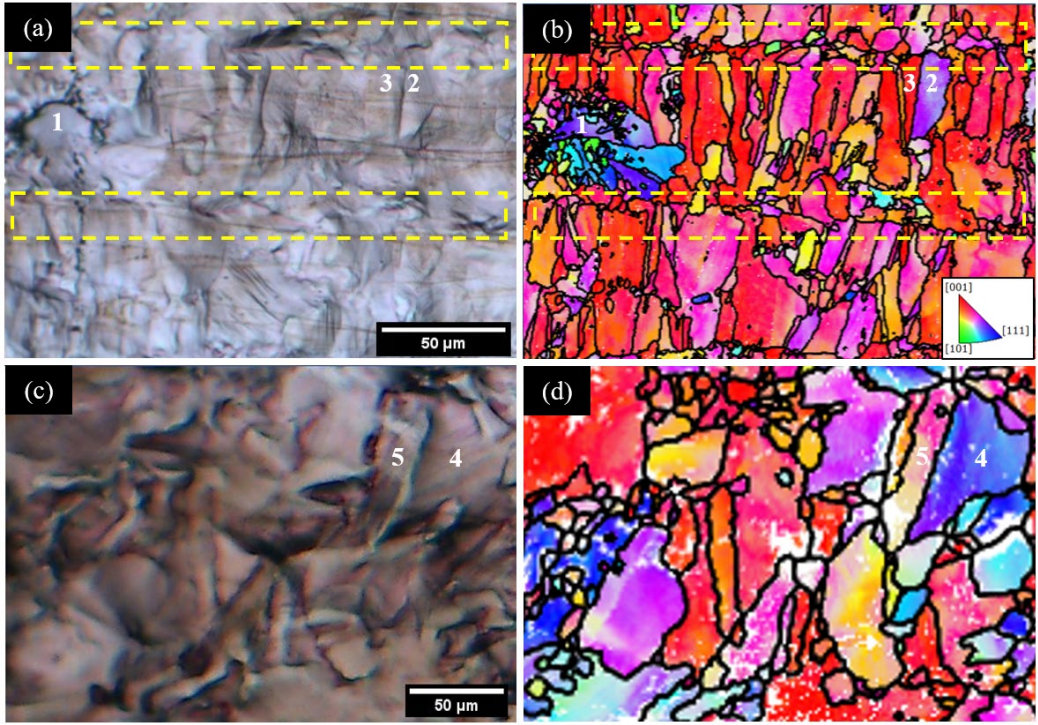

Figure 7. Influence of grain boundaries on strain localisation in the B0 condition (top row) and the CB67 condition (bottom row). (a,c) Optical micrographs and (b,d) corresponding EBSD-IPF along the loading direction. Some grains of significant misorientation across the grain boundaries were numbered to highlight locations where there were considerable changes in surface roughness. The effect of line rows of fine grains on localisation in the B0 condition is evident in (a).

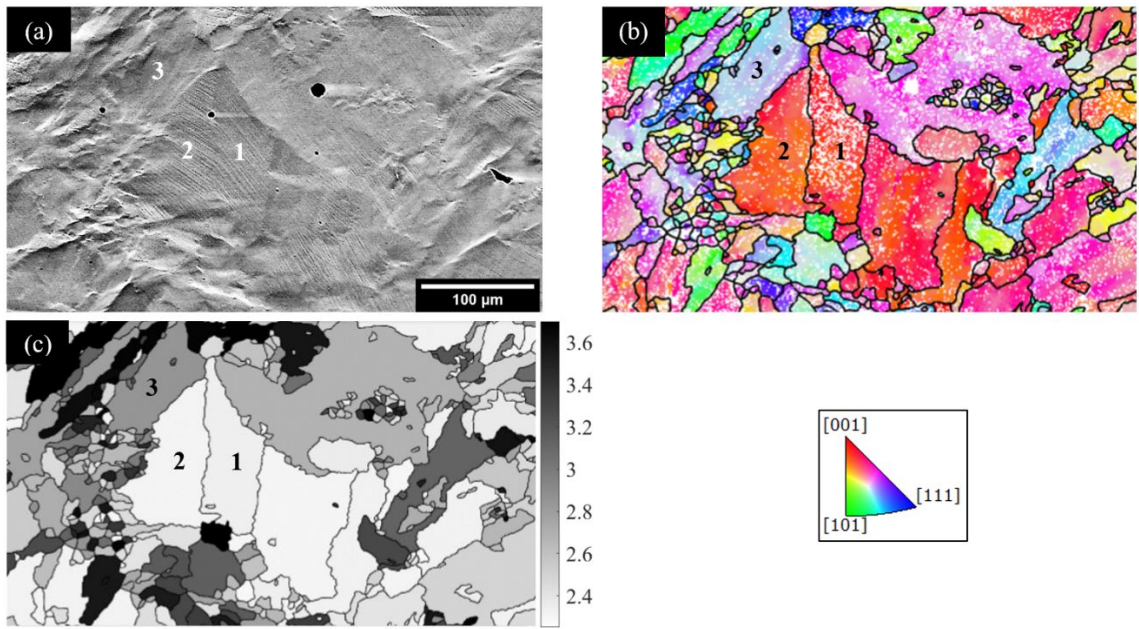

Figure 8. Slip transmission across neighbouring grains in the CB67 condition at 10% strain. All figures are of the same region: (a) SEM image; (b) EBSD-IPF along the loading direction; (c) Taylor factor map.

Figure 9 shows IPF scans along the LD with the corresponding GND maps in the gauge region near the fracture surface. On average, the GND density substantially increased from the as-built and undeformed condition of the B0 (and CB67) samples, 7.6 x $10^{13}$ m$^{-2}$ (and 6.1 x $10^{13}$ m$^{-2}$) to 1.9 x $10^{14}$ m$^{-2}$ (and 1.5 x $10^{14}$ m$^{-2}$) after fracture, respectively. Most notably, the GND accumulation increased within both the soft and most of hard grains, highlighting that the hard grains were eventually plastically deformed. The location of high GND accumulation appeared to be along the cell boundaries, the same as it was before deformation. This suggests a significant role of the AM-inherent cellular structures in impeding and retaining dislocations in AM-processed alloys, contributing to increasing the strength of the material and regulating the spatial plastic deformation. While in wrought alloys dislocation accumulation occurs most intensely near high-angle grain boundaries, in AM-processed alloys the boundaries of cellular structures within grains serve as additional effective obstacles to dislocation motion, hence accumulation sites of dislocations.

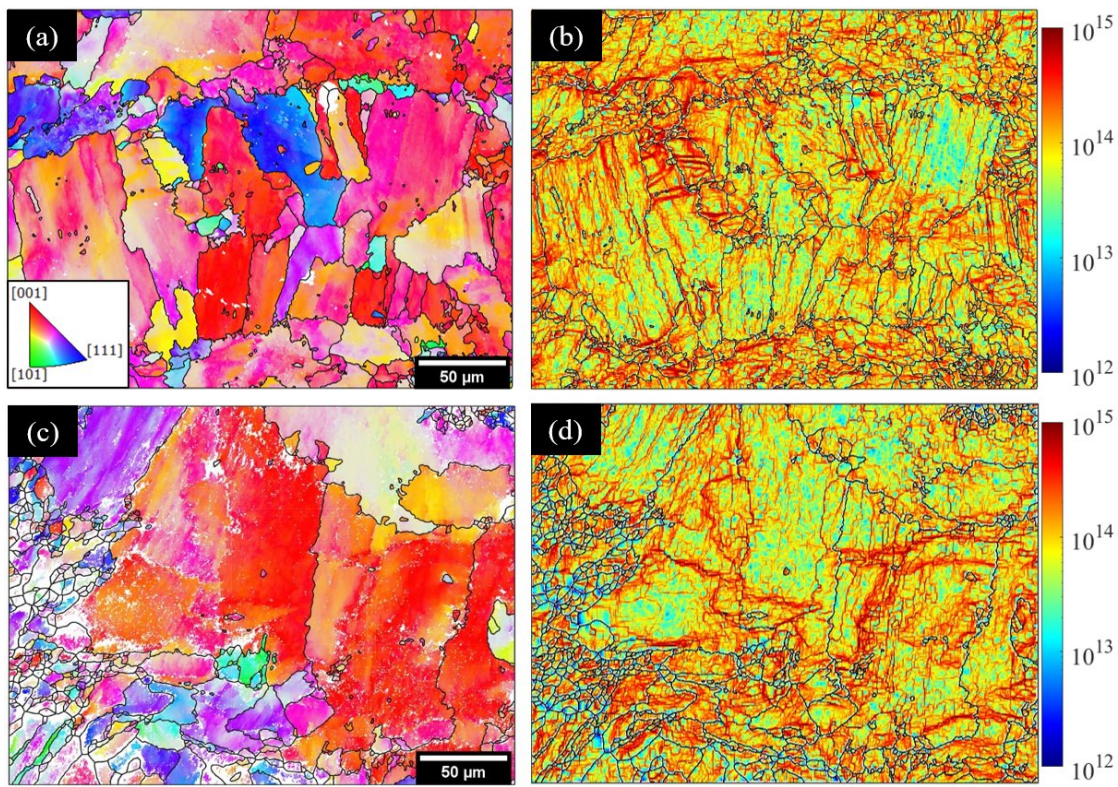

Figure 9. EBSD-IPF along the loading direction with their corresponding GND maps in the gauge region near the fracture surface: (a) and (b) B0 condition; (c) and (d) CB67 condition.

Figure 10(a-h) shows the SEM and EBSD IPF (along the loading direction) in the undeformed condition and at 10% strain during the *in-situ* test for the B0 and CB67 samples. In agreement with the observation given in Figure 7, in comparison to the CB67 the surface of the B0 condition varied much more significantly and more intense localisation developed in the <001> || LD grains after the nominal 10% strain, Figure 10(c-d) versus (g-h). The surface roughening and build-up of dislocations with the increasing strain intervals led to degrading the EBSD indexing, which was much worse in the B0 sample.

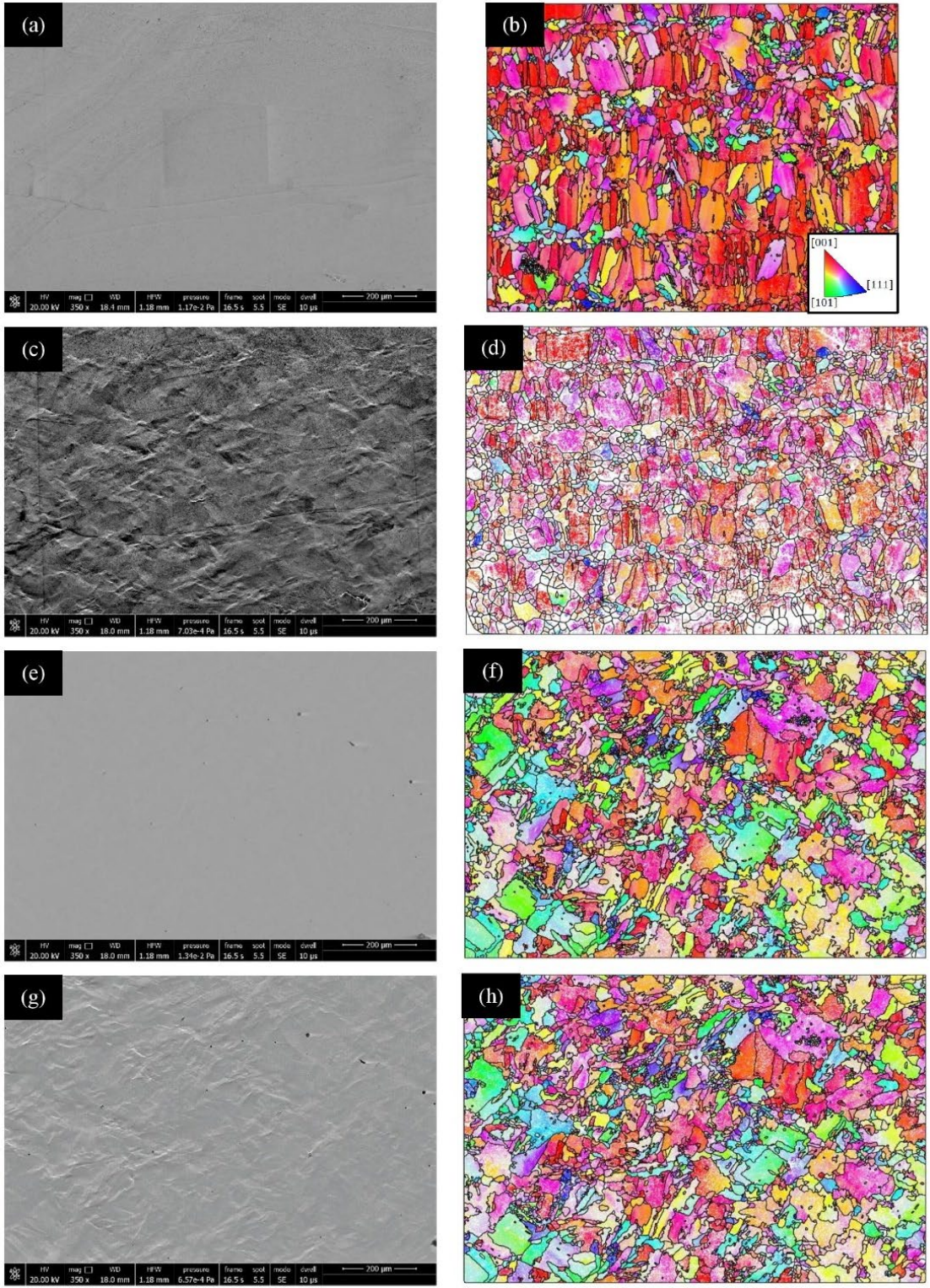

Figure 10. SEM images and their corresponding IPF maps along the loading direction for (a-b) undeformed B0 condition; (b-c) deformed B0 condition at 10% strain; (e-f) undeformed CB67 condition; (g-h) deformed CB67 condition at 10% strain. All figures have the same scale bar.

### 3.3. Texture evolution during plastic deformation

Crystallographic texture is an influential factor in the plastic deformation behaviour spatially and globally [47]. AM alloys have been often reported to be strongly textured in the as-built condition. Therefore, it is necessary to study the evolution of texture components with strain. Main texture components such as Cube, Copper, S, Brass and Goss (Table 1) were quantified from the as-built (undeformed) condition to fracture for the B0 and CB67 samples, Figure 11. The undeformed B0 microstructure contained mostly the Cube (35.8%) and Goss (27.0%) texture components. During the *in-situ* tensile test, the Cube and Goss components in the B0 microstructure most rotated at the lower strain range. For example, after a nominal strain of 4%, the Cube weakened by 3.1% while the Goss increased by 2.7%. Such changes were more significant than those of other texture components such as Copper (decrease by 0.4%), S (increase by 2.0%) and Brass (increase by 1.4%), Figure 11a. The Cube orientations belong to the soft <001> fibre and have a low Taylor factor (and similar for the Goss orientations if the loading is parallel to a <001>), whereas the Copper and Brass on the <111> fibre are not preferably oriented for plasticity and are more resistant to deformation and grain rotation (hence texture changes), as discussed in section 3.2. The stability of the Brass component in Ni-based superalloys is in agreement with previous studies [46,48,49]. Figure 11b reveals that at the point of fracture, the Cube texture decayed while the Goss and Brass components intensified in the B0 condition.

Figure 11c shows that the Cube texture component was also the most dominant in the as-built CB67 condition, but lower than in the initial B0 condition. Also, contrary to the B0, the Goss component was comparatively much lower in the initial CB67 condition. While the Cube and Goss components showed minimal change throughout the deformation of the CB67 (Figure 11d), the Copper, S and Brass components reorientated and most notably changed at strain levels exceeding 10%, with the Copper component intensifying the most and increasing by 6.9% by the point of fracture. The difference in the grain reorientation behaviour is governed by both the initial crystallographic orientations of the individual grains as well as the neighbouring grains. It was theorised by Raabe *et al.* [49] that the deformation and rotation paths of the Goss and Cube grains, in particular, were strongly influenced by the neighbouring grains. This is also evident in the different evolution of texture components in the B0 and CB67 conditions in this study.

It has been previously demonstrated that the Goss texture component causes the most plastic anisotropy in components, whereas the Cube component results in the least anisotropy (apart from the totally random texture) [30]. This study shows that the as-built B0 condition contained a higher fraction of the preferable Cube component than that in the CB67 condition. However, the Cube component degraded quickly with plastic deformation in the B0 condition, but the Goss texture component, which was initially also high in the as-built B0 condition, increased even more during loading. The decrease in Cube and the increase in Goss in the B0 condition during deformation could intensify the plastic anisotropy. On the contrary, the dominant Cube texture and the substantially lower fraction of the Goss component remained stable throughout the plastic deformation of the CB67 condition, helping to maintain a homogeneous deformation behaviour.

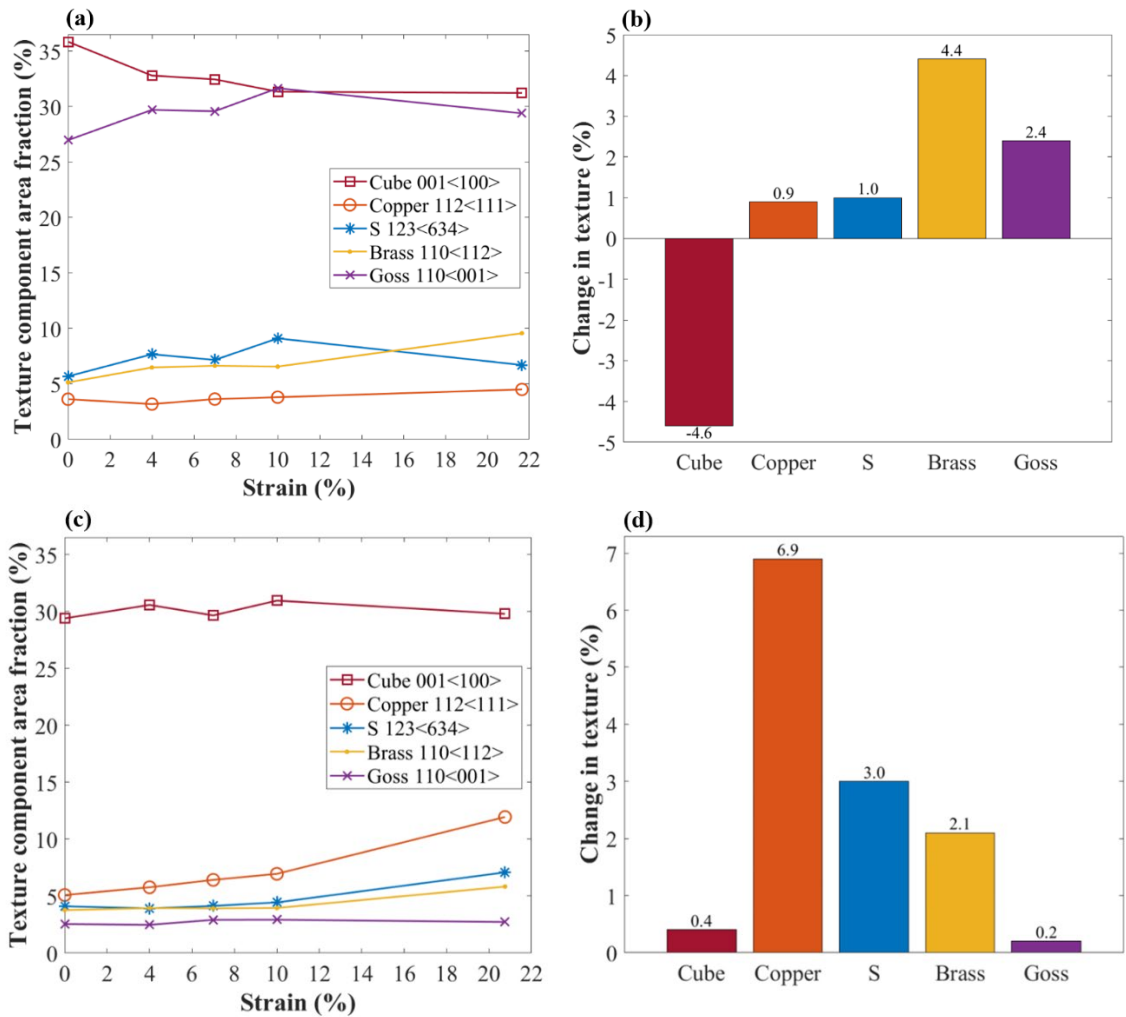

Figure 11. Evolution of texture component area fraction across the strain intervals and their overall change from the as-built to fractured: (a-b) B0 condition and (c-d) CB67 condition.

### 3.4. Impact of scanning strategy on plastic inhomogeneity

Plastic inhomogeneity in AM alloys is governed by the arrangement, morphology, size distribution, and crystallographic texture of grains and the spatial distribution of dislocations. The results presented in sections 3.1 and 3.3 demonstrate that such microstructure can be effectively controlled by the laser scanning strategy, thereby influencing the spatial plastic deformation. The B0 strategy creates elongated columnar grains of alternating <001> and <101> orientations with respect to the building direction (Figure 3a). The top view perpendicular to the building direction of the B0 condition showed highly ordered alternating rows of fine grains of Cube <001> orientations (containing a very high dislocation density, including GNDs) and coarse grains of Goss <101> orientations (with a lower density of GNDs that are aligned perpendicular to the scan direction, Figure 3b-d). In essence, the B0 strategy creates segregated zones of two dominant textures having significantly different deformation behaviour, inducing intense strain localisation and severe anisotropy under mechanical loading. In addition to the aligned distribution of GNDs, such an arrangement of texture components significantly contributes to the strain localisation and surface topography variation observed in the B0 condition. In contrast, the island scanning and interlayer rotation in the CB67 strategy interrupt the epitaxial columnar growth and result in more random grain morphologies, crystallographic orientations, and GND distribution (Figure 3e-f). Although the preferred Cube texture component area fraction was shown to be the most dominant component in both the B0 and CB67 conditions, the CB67 contained a much lower fraction of the Goss component, which is the most anisotropic component [30]. In addition, the Cube and Goss texture components were found to remain stable throughout the plastic deformation of the CB67 condition (Figure 11c-d), retaining the plastic homogeneity and minimising plastic localisation even after plastic straining (Figure 10g-h) – this is desirable for AM builds.

### 4. Conclusions

This work investigates the spatial aspects of key microstructure such as the grain arrangement, dislocation condition (density and distribution) and crystallographic texture (namely Cube, Goss, Copper, Brass and S components), and their effects on the plastic deformation inhomogeneity in Inconel 718 fabricated by laser powder-bed fusion with varied scan strategies. *In-situ* tensile testing coupled with optical microscopy, SEM and EBSD was also performed to examine the evolution of such key microstructure and how they affect the spatial deformation behaviour.

The bidirectional strategy without layer rotation (i.e. B0) resulted in a highly ordered arrangement of coarse grains separated by line clusters of fine grains along the centre of deposition tracks. The fine grains contained a much higher density of dislocations, in particular geometrically necessary dislocations (GNDs), than the coarse grains. The GND alignment in the coarse grains was polarised and primarily orthogonal to the laser scanning direction due to the fast laser speed used. The fine grains were found to have strong Cube texture (which induces good isotropy) while the coarse grains were of strong Goss textures (which often causes severe anisotropy). Under the uniaxial tension, the Cube was weakened while the Goss was strengthened. Such microstructure arrangement and distribution promoted plastic inhomogeneity, in particular strain localisation. It was found that (1) plastic strain strongly localised in the line cluster of the fine grains, and (2) surface topography significantly changed particularly at grain boundaries between grains of Cube texture and Goss texture. As plastic strain localisation is a main mechanism responsible for crack initiation, such microstructure is unfavourable for high-performance builds made by additive manufacturing. In contrast, the microstructure created by the chessboard strategy with 67° layer rotation (i.e. CB67) was much more favourable. In particular, texture and dislocation distribution were much more random, dislocation density was lower, and the orderly alternating arrangement of coarse/fine grains found in the B0 condition was eliminated. The CB67 strategy also resulted in strong Cube, but much weaker Goss component. The Cube component was stable during plastic deformation, making it much more favourable for homogeneous deformation.

**Acknowledgements**

J. Al-Lami would like to thank Richard Cross, Edward Cross and Aidan Bew at Cross Manufacturing Ltd for supporting this work.